\documentclass[english,aps,graphicx,twocolumn]{revtex4}
\usepackage[T1]{fontenc}
\usepackage[latin9]{inputenc}
\usepackage{color}
\usepackage{array}
\usepackage{booktabs}
\usepackage{amsmath}
\usepackage{amssymb}
\usepackage{graphicx}
\usepackage{esint}
\usepackage{float}
\usepackage{babel}
\makeatletter

\newcommand{\lyxmathsym}[1]{\ifmmode\begingroup\def\b@ld{bold}
  \text{\ifx\math@version\b@ld\bfseries\fi#1}\endgroup\else#1\fi}

\providecommand{\tabularnewline}{\\}

\makeatother

\begin{document}
%\begin{CJK*}{GBK}{song}%\begin{CJK*}{GBK}{song} %ÏÔÊŸÖÐÎÄ
\preprint{APS/123-QED}
\title{Quantum metrology with one auxiliary particle in a correlated bath \\ and its quantum simulation}

\author{Wan-Ting He,$^{1,*}$
Huan-Yu Guang,$^{1,*}$
Zi-Yun Li,$^{1,}$\footnote{These authors contributed equally to this work.}
Ru-Qiong Deng,$^{2,1}$\\
Na-Na Zhang,$^{1}$
Jie-Xing Zhao,$^{1}$\footnote{zhaojiexing@mail.bnu.edu.cn}
Fu-Guo Deng,$^{1}$
and Qing Ai,$^{1,}$\footnote{aiqing@bnu.edu.cn}}

\address{Department of Physics, Applied Optics Beijing Area Major Laboratory,
Beijing Normal University, Beijing 100875, China\\
$^{2}$School of Physics and Electronics, Hunan Normal University, Hunan 410006, China
}

%\author{Wan-Ting He}
%\affiliation{Department of Physics, Applied Optics Beijing Area Major Laboratory,
%Beijing Normal University, Beijing 100875, China}
%\affiliation{These authors contribute equally to this work.}
%
%\author{Huan-Yu Guang}
%\affiliation{Department of Physics, Applied Optics Beijing Area Major Laboratory,
%Beijing Normal University, Beijing 100875, China}
%\affiliation{These authors contribute equally to this work.}
%
%\author{Zi-Yun Li}
%\affiliation{Department of Physics, Applied Optics Beijing Area Major Laboratory,
%Beijing Normal University, Beijing 100875, China}
%\affiliation{These authors contribute equally to this work.}
%
%\author{Ru-Qiong Deng}
%\affiliation{Hunan Normal University, Hunan 410006, China.}
%
%\author{Jie-Xing Zhao}
%\email{zhaojiexing123@163.com}
%\affiliation{Department of Physics, Applied Optics Beijing Area Major Laboratory,
%Beijing Normal University, Beijing 100875, China}
%
%\author{Qing Ai}
%\email{aiqing@bnu.edu.cn}
%\affiliation{Department of Physics, Applied Optics Beijing Area Major Laboratory,
%Beijing Normal University, Beijing 100875, China}

\date{\today}
\begin{abstract}
In realistic metrology, entangled probes are more sensitive to noise,
especially for a correlated environment. The precision of parameter estimation
with entangled probes is even lower than that of the unentangled ones
in a correlated environment. In this paper, we propose a measurement scheme with only one auxiliary qubit, which can selectively offset the impact of environmental noise
under this situation. We analyse the estimation precision of our scheme
and find out that it approaches the Heisenberg limit when prepared in a proper
auxiliary state. We further discuss employing auxiliary states to improve
the precision of measurement in other environment models such as a
partially-correlated environment. In order to verify our scheme, we apply a recently-developed quantum algorithm to simulate the quantum dynamics of our proposal and show that it outperform the other proposals with less resources.
\end{abstract}
\maketitle

%\end{CJK*}
\section{introduction}
\label{intro:}
Quantum metrology employs quantum entanglement and coherence to achieve an ultra-high
precision for the estimation of an unknown parameter \citep{Advanced,2017QuantumSensing}. It has become an indispensable element of satellite navigation, aerospace measurement
and control, mobile phone and computer chip processing. This opens
a broad range of valuable applications of quantum mechanics, in addition to quantum information processing \citep{QuantumComputation}, photosynthetic exciton energy transfer \citep{2013Biology,2013Clustered}, avian magnetoreception \citep{2012Sensitive,2012Generalized,2013Biology,2013ChemicalCompass} and quantum metamaterial \citep{2004Electromagnetically,2006Electromagnetically,2007Tunable,Hyperbolic2020}. As a result of the central-limit theorem, the estimated precision of an unknown quantity critically depends on the number of resources available for the measurement. One of the primary goals of quantum metrology is to enhance the precision of resolution with limited resources.
An enhanced resolution can be achieved
if quantum entanglement is used to correlate the probes before making them interact with the system to be measured \citep{1992Quantum,Giovannetti2004Quantum}.
Taking advantage of $n$ entangled quantum probes,
one can attain the Heisenberg limit (HL) which scales as $n^{-1}$,
being the ultimate limit in precision set by quantum mechanics. And this result has been demonstrated experimentally \citep{Leibfried2004Toward,Roos2006}.
On the contrary, when using unentangled probes, one can only reach the standard quantum
limit (SQL) which scales as $n^{-1/2}$. Obviously, the use of entanglement
can significantly enhance the precision when $n$ is large.

In realistic scenario for experiments, quantum probes are inevitably
affected by noise. The achievable precision decreases due to the decoherence.
As known to all, the quantum dynamics of open quantum systems are usually classified into Markovian and non-Markovian \cite{2016Colloquium,2017Dynamics,2018Concepts,2009Degree,2011ExperimentalControl,2010Entanglement}. When the system-bath couplings are relatively large, or the number of degrees of freedom in the environment is not sufficiently large, e.g.
natural photosynthetic complexes and NV centers in diamond, the open quantum systems are subject to non-Markovian quantum dynamics \citep{2013Biology,2006QuantumLimits}.
Using entangled probes in a non-Markovian quantum dynamics allows for a higher
measurement precision than that in a Markovian quantum dynamics, which scales as $n^{-3/4}$ \citep{2012Quantum}. On the other hand, when particles interact with a correlated environment, e.g. nuclear spins in a molecule, the decoherence rate per particle will increase linearly with the number of particles,
i.e., superdecoherence \citep{2019Conditions}. In this regime, utilizing entangled probes will no longer outperform
unentangled ones no matter the open quantum dynamics is Markovian or non-Markovian.

In order to overcome decoherence in an uncorrelated bath, logical states are introduced to establish decoherence-free subspace \citep{Dorner2012Quantum,2017Enhancement,Song2016}.
However, since $\mathit{N}$ logical qubits require $n=2N$ physical qubits,
it doubles the number of valuable resources used per experiment.
In order to effectively reduce the usage of resources,
here we propose a measurement scheme with only one auxiliary qubit in a correlated bath. Using such a well-designed auxiliary particle in quantum metrology can selectively offset the impact of noise in a correlated environment. Comparing with previous proposals, we show that although we use less resources, we can still approach the HL when preparing
proper auxiliary states.

On the other hand, although the theoretical scheme utilizing entanglement in a non-Markovian environment is appealing, it might be difficult to experimentally verify it since it critically requires the homogeneity of the qubits. For example, for NV centers in diamond, different spins manifest different Zeeman energies and decoherence rates in the same environment due to the inhomogeneous gyromagnetic ratios. Recently, based on the bath-engineering technique \citep{2014ExperimentalNoise,2014Soare} and the gradient ascent pulse engineering (GRAPE) algorithm \citep{2005OptimalControl,2017Hybrid}, we theoretically proposed and experimentally demonstrated that the open quantum dynamics with an arbitrary Hamiltonian and spectral density can be exactly and efficiently simulated \cite{WangEfficient,Nana2020}. Thus, based on this algorithm, we show how to verify that our scheme can make the estimation precision achieve the HL in a quantum simulation experiment \citep{Simulation1,Simulation2}.

This paper is organized as follows: Our measurement scheme and its quantum dynamics for quantum metrology in a correlated bath are introduced in the next section. Specifically, we offer an example of magnetic-field sensing utilizing our measurement scheme. Then, in Sec.~\ref{Sec2}, we apply a recently-developed quantum algorithm to simulate our measurement scheme with an auxiliary qubit and show that it can approach the HL.
In Appendix, we give a brief description for the open quantum dynamics of $n$ qubits in a correlated bath and an uncorrelated bath, respectively.

\section{Model and Dynamics}
\label{main section}

\subsection{Quantum Metrology under Superdecoherence}
\label{subsection2}

In quantum metrology, using entangled probes can obtain an increase
in precision when assuming a fully-coherent evolution, while in realistic
scenario, there is always decoherence caused by
environmental noise. The optimal precision with entangled probes under
decoherence was first analyzed in Markovian environment~\citep{97PRL}, and further
discussed in non-Markovian environment~\citep{2012Quantum}. The best resolution in the
estimation was obtained with a given number of particles
$n$ and a total duration of experiment $T$, i.e., $nT/t$ being
the actual number of experiment. The entangled probes are prepared
in an $n$-qubit GHZ state $(|0\rangle^{\otimes N}+|1\rangle^{\otimes N})/\sqrt{2}$,
and then evolve freely for a duration $t$. In the ideal case, we have $(|0\rangle^{\otimes N}+\exp(-in\phi t)|1\rangle^{\otimes N})/\sqrt{2}$, where $\phi$ is the parameter to be estimated. However, in the presence
of dephasing noise, the final state becomes $\{|0\rangle^{\otimes N}+\exp[-in\phi t-\Gamma_{n}(t)]|1\rangle^{\otimes N}\}/\sqrt{2}$, with $\Gamma_{n}\left(t\right)=\int_{0}^{t}d\tau\gamma_{n}(\tau)$,
and $\gamma_{n}(t)$ being the dephasing rate of $n$ qubits. After a $\pi/2$ pulse, the probability
of finding these probes in the initial state reads
\begin{equation}
P=\frac{1}{2}\left[1+\cos(n\phi t)e^{-\Gamma_n(t)}\right].
\end{equation}
The uncertainty of the measurement can be calculated as
\begin{equation}
\delta\phi^{2}=\frac{1}{\left(nT/t\right)F\left(\phi\right)},
\end{equation}
where the Fisher information \citep{OpenQuantumSystems,LiuP2017} is
\begin{equation}
F(\phi)\equiv\frac{1}{P(1-P)}\left(\frac{\partial P}{\partial\phi}\right)^{2}=\frac{t^{2}\sin^{2}(\phi t)e^{-2\Gamma_n(t)}}{1-\cos^{2}(\phi t)e^{-2\Gamma_n(t)}}.
\end{equation}
When the entangled probes are in uncorrelated environment, we can
obtain $\Gamma_{n}(t)=n\int_{0}^{t}d\tau\gamma(\tau)=n\Gamma(t)$,
where $\gamma(t)$ and $\Gamma(t)$ are respectively the dephasing rate and decoherence factor for a single qubit.
Thus, the uncertainty is explicitly written as \citep{97PRL,2012Quantum}
\begin{equation}
\delta\phi^{2}=\frac{1-\cos^{2}\left(\phi t\right)e^{-2n\Gamma\left(t\right)}}{nTt\sin^{2}\left(\phi t\right)e^{-2n\Gamma\left(t\right)}}.
\end{equation}
In order to attain the best precision, it is necessary to optimize
this expression of uncertainty against the duration of each
single measurement $t$. The best interrogation time satisfies $\phi t_{e}=k\pi/2$
with odd $k$ and $2nt\frac{d\Gamma(t)}{dt}\mid_{t=t_{e}}=1$ \citep{2012Quantum},
and thus yields
\begin{equation}
\delta\phi^{2}\mid_{e}=\frac{1}{n^{2}Tt_{e}}e^{2n\Gamma\left(t_{e}\right)},
\end{equation}
where the subscript $e$ indicates that the entangled probes are used.

When the entangled probes are in a correlated environment, the superdecoherence
of the probes will modify the probability as
\begin{equation}
P=\frac{1}{2}\left[1+\cos\left(n\phi t\right)e^{-n^{2}\Gamma\left(t\right)}\right].\label{eq:probability}
\end{equation}
The uncertainty of parameter $\phi$ reads
\begin{equation}
\delta\phi^{2}|_{e}=\frac{1-\cos^{2}\left(n\phi t\right)e^{-2n^{2}\Gamma\left(t\right)}}{nTt\sin^{2}\left(n\phi t\right)e^{-2n^{2}\Gamma\left(t\right)}}.
\end{equation}
By minimizing $\delta\phi^{2}|_{e}$, i.e., requiring that $\phi t_{e}=k\pi/2$ with odd $k$ and $2n^{2}t\frac{d\Gamma\left(t\right)}{dt}\mid_{t=t_{e}}=1$, we have
\begin{equation}
\delta\phi^{2}|_{e}=\frac{1}{n^{2}Tt_{e}}e^{2n^{2}\Gamma\left(t_{e}\right)}.\label{eq:deltaPhiE}
\end{equation}

When utilizing unentangled probes, the uncertainty of parameter is $\delta\phi^{2}\mid_{u}=\frac{1}{nTt_{u}}\exp[2\Gamma(t_{e})]$,
and the best interrogation time is given by $\phi t_{u}=k\pi/2$
with odd $k$ and $2t\frac{d\Gamma(t)}{dt}\mid_{t=t_{u}}=1$
\citep{2012Quantum}. Following Ref.~\citep{2012Quantum}, we define
\begin{equation}
r=\frac{\delta\phi|_{u}}{\delta\phi|_{e}},
\end{equation}
where $\left|\delta\phi\right|_{e}$ and $\left|\delta\phi\right|_{u}$
are the standard deviation of the estimated parameter $\phi$ when
using entangled and unentangled probes, respectively. Here, $r$ characterizes the improved precision of measurement for using entangled probes instead of unentangled ones. For entangled
probes in an uncorrelated environment, we obtain $r^{2}=n(t_{e}/t_{u})\exp[2\Gamma(t_{u})-2n\Gamma(t_{e})]$,
while for entangled probes in a correlated environment, we have $r^{2}=n(t_{e}/t_{u})\exp[2\Gamma(t_{u})-2n^2\Gamma(t_{e})]$.
Obviously, $r$ changes along with the dependence of function $\Gamma(t)$ on time \citep{2012Quantum},
i.e., the dynamics of decoherence. For example, $\Gamma(t)\propto t$ corresponds
to the Markovian dephasing dynamics. When $\Gamma(t)$
has a quadratic behavior, i.e., $\Gamma(t)\propto t^{2}$, it is
the non-Markovian dephasing dynamics, which is actually the quantum Zeno dynamics \cite{2017Enhancement,2010Quantum,Ai2013SR,Harrington2017}. Hereafter, we just follow the terminology in Ref.~\citep{2012Quantum}. In Tab.~\ref{tab:table1}, we analyze the relative resolution of the parameter
$r$ in different situations, i.e., Markovian vs non-Markovian
dynamics and correlated vs uncorrelated environment.
We can learn from Tab.~\ref{tab:table1} that
using entangled probes can not improve the precision in the presence of superdecoherence because $r\leq1$ for $n\geq2$.

\begin{table}
\caption{\label{tab:table1}The relative parameter resolution $r$ varies with
different dephasing dynamics}

\begin{tabular}{>{\centering}p{75pt}>{\centering}p{75pt}>{\centering}p{75pt}}
\toprule
 & {\footnotesize{}uncorrelated environment} & {\footnotesize{}correlated environment}\tabularnewline
\midrule
\midrule
{\footnotesize{}Markovian} & {\footnotesize{}$r=1$} & {\footnotesize{}$r=n^{-1/2}$}\tabularnewline
\midrule
{\footnotesize{}non-Markovian} & {\footnotesize{}$r=n^{1/2}$} & {\footnotesize{}$r=1$}\tabularnewline
\bottomrule
\end{tabular}
\end{table}

\subsection{Quantum Metrology Using an Auxiliary Particle}
\label{subsection3}
As shown in the pervious section, on account of superdecoherence, the best precision with entangled
probes will no longer be superior to that with unentangled ones.
Previous works \citep{Dorner2012Quantum,2017Enhancement} introduced
logical states which are decoherence-free to improve the precision
at the cost of doubled resources used per experiment.
In this section, we present a measurement scheme with auxiliary states,
only one auxiliary qubit being used per experiment. When we prepare a
properly-designed auxiliary qubit, the precision of parameter
estimation can approach the HL.

\begin{figure}
\includegraphics[bb=0 0 200 118,width=8.5cm]{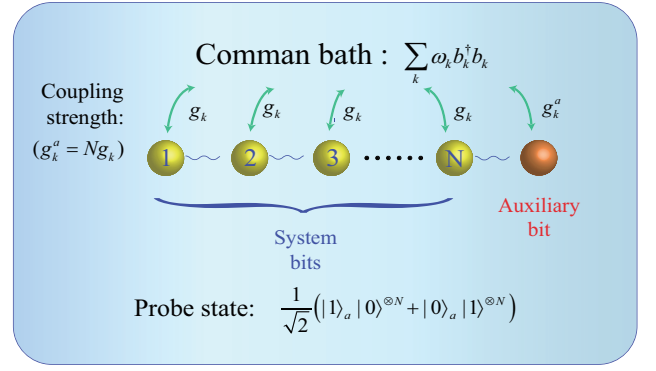}
\caption{Schematic for $N$-qubits and 1 auxiliary qubit in a correlated bath.
An $(N+1)$-qubit entangled system, initially at $|\Psi(0)\rangle=(|1\rangle_{a}|0\rangle^{\otimes N}+|0\rangle_{a}|1\rangle^{\otimes N})/\sqrt{2}$, interacts with a correlated bath.
The coupling constants between the qubits and the $k$th
mode of the bath satisfy $g_{k}^{a}=Ng_{k}$.}\label{fig:Schematic-illustration}
\end{figure}

As illustrated in Fig.~\ref{fig:Schematic-illustration}, we initially prepare an $(N+1)$-qubit entangled state with $N$ being the number of working qubits, i.e.,
\begin{equation}
|\Psi(0)\rangle=\frac{1}{\sqrt{2}}\left(|1\rangle_{a}|0\rangle^{\otimes N}+|0\rangle_{a}|1\rangle^{\otimes N}\right),\label{eq:initial state}
\end{equation}
where the subscript $a$ indicates the auxiliary qubit.

The total system is governed by the Hamiltonian $H=H_\textrm{S}+H_\textrm{B}+H_\textrm{int}$,
$H_\textrm{S}$ and $H_\textrm{B}$ being the Hamiltonian of probe and bath, $H_\textrm{int}$
being the interaction Hamiltonian
\begin{eqnarray}
H&=& \frac{1}{2}\Omega_{0}\sum_{i=1}^{N}\sigma_{z}^{(i)}+\frac{1}{2}\omega_{a}\sigma_{z}^{a}+\sum_{k}\omega_{k}b_{k}^{\dagger}b_{k}\nonumber\\
 &  & +\sum_{k}\sum_{i=1}^{N}(g_{k}^{(i)}\sigma_{z}^{(i)}+g_{k}^{a}\sigma_{z}^{a})(b_{k}^{\dagger}+b_{k}),\label{eq:Model Hamiltonian}
\end{eqnarray}
where $\omega_{a}$ is the frequency of the auxiliary qubit, we assume identical frequency $\Omega_{0}$
for the $N$ working qubits, $\sigma_{z}^{(i)}$
and $\sigma_{z}^{a}$ are the Pauli operators of the working and auxiliary
qubits, respectively,
$g_{k}$ ($g_{k}^{a}$) denotes the coupling constant between the working (auxiliary) qubit
and the $k$th mode of the environment,
$b_k^\dagger$ is the creation operator of the harmonic oscillator with frequency $\omega_k$.

According to Eq.~(\ref{eq:BR new}), the quantum dynamics of the reduced density matrix of the total system including the working and auxiliary qubits is described by the quantum master equation
\begin{eqnarray}
\dot{\rho}&=& i[\rho,H_\textrm{S}+H_\textrm{LS}]\nonumber \\
 &  & +\frac{1}{2}[\sum_{i,j=1}^{N}C_{ij}(0,t)(\sigma_{z}^{(j)}\rho\sigma_{z}^{(i)}-\frac{1}{2}\{ \sigma_{z}^{(i)}\sigma_{z}^{(j)},\rho\} )\nonumber \\
 &  & +\sum_{i=1}^{N}C_{ia}(0,t)(\sigma_{z}^{a}\rho\sigma_{z}^{(i)}-\frac{1}{2}\{ \sigma_{z}^{(i)}\sigma_{z}^{a},\rho\} )\nonumber \\
 &  & +\sum_{i=1}^{N}C_{ai}(0,t)(\sigma_{z}^{(i)}\rho\sigma_{z}^{a}-\frac{1}{2}\{ \sigma_{z}^{a}\sigma_{z}^{(i)},\rho\} )\nonumber \\
 &  & +C_{aa}(0,t)(\sigma_{z}^{a}\rho\sigma_{z}^{a}-\rho)],\label{eq:SA equation}
\end{eqnarray}
where the time correlation functions are
\begin{eqnarray}
C_{AB}(\omega,t)=2\sum_{k}g_{k}^{A}g_{k}^{B}\left[2\overline{n}(\omega_{k},T)+1\right]\frac{\sin\omega_{k}t}{\omega_{k}}|_{\omega_k=\omega}
\end{eqnarray}
for $A,B=i,j,a$ with $\overline{n}(\omega_{k},T)$ being the Bose-Einstein distribution at temperature $T$, $\{A,\rho\}=A\rho+\rho A$ is the anti-commutator,
\begin{eqnarray}
H_\textrm{LS}&=& \sum_{i,j=1}^{N}F_{ij}(0,t)\sigma_{z}^{(i)}\sigma_{z}^{(j)}+\sum_{i=1}^{N}F_{ia}(0,t)\sigma_{z}^{(i)}\sigma_{z}^{a}\nonumber \\
 &  & +\sum_{i=1}^{N}F_{ai}(0,t)\sigma_{z}^{a}\sigma_{z}^{(i)}+F_{aa}(0,t)\mathbb{I},
\end{eqnarray}
is the Hamiltonian with the Lamb shift
\begin{eqnarray}
F_{AB}(\omega,t)=\sum_{k}g_{k}^{A}g_{k}^{B}\frac{\cos\omega_{k}t-1}{\omega_{k}}|_{\omega_k=\omega}
\end{eqnarray}
for $A,B=i,j,a$, $\mathbb{I}$ is the identity operator.

\subsubsection{Correlated Environment}

Suppose entangled probes are physically close, i.e., $g_k^{(j)}=g_k$ for $j=1,2,\cdots N$.
They may suffer from superdecoherence with the time correlation function $C_{ij}(0,t)=\gamma(t)$ for $i,j=1,2,\cdots N$.
As illustrated in Fig.~\ref{fig:Schematic-illustration}, in our
measurement scheme, we prepare a proper auxiliary qubit, whose coupling
constant with the $k$th mode in the bath reads $g_{k}^{a}=Ng_{k}$.
Thus, we have $C_{ia}(0,t)=C_{ai}(0,t)=N\gamma(t)$, and $C_{aa}=N^{2}\gamma(t)$.
Equation~(\ref{eq:SA equation}) is rewritten as
\begin{eqnarray}
\dot{\rho}&=& i[\rho,H_\textrm{S}+H_\textrm{LS}]\nonumber \\
 &  & +\frac{1}{2}[\sum_{i,j=1}^{N}\gamma(t)(\sigma_{z}^{(j)}\rho\sigma_{z}^{(i)}-\frac{1}{2}\{ \sigma_{z}^{i}\sigma_{z}^{(j)},\rho\} )\nonumber \\
 &  & +N\sum_{i=1}^{N}\gamma(t)(\sigma_{z}^{a}\rho\sigma_{z}^{(i)}-\frac{1}{2}\{ \sigma_{z}^{(i)}\sigma_{z}^{a},\rho\} )\nonumber \\
 &  & +N\sum_{i=1}^{N}\gamma(t)(\sigma_{z}^{(i)}\rho\sigma_{z}^{a}-\frac{1}{2}\{ \sigma_{z}^{a}\sigma_{z}^{(i)},\rho\} )\nonumber \\
 &  & +N^{2}\gamma(t)(\sigma_{z}^{a}\rho\sigma_{z}^{a}-\rho)].
\end{eqnarray}
For simplicity, we denote $|1\rangle\equiv|0_{a}\rangle|1\rangle^{\otimes N}$ and
$|2\rangle\equiv|1_{a}\rangle|0\rangle^{\otimes N}$. The diagonal terms of the density matrix are constant in time,
%of the off-diagonal element of the density matrix
%\begin{eqnarray}
%\dot{\rho}_{1,2}(t)= &  & \langle1|\{i\left[\rho,\frac{1}{2}\omega_{0}\sum_{i=1}^{N}\sigma_{z}^{i}+\frac{1}{2}\omega_{a}\sigma_{z}^{a}\right]\nonumber \\
% &  & +i\left[\rho,H_{LS}\right]\nonumber \\
% &  & +\frac{1}{2}[\sum_{i,j=1}^{N}\gamma(t)\left(\sigma_{z}^{j}\rho\sigma_{z}^{i}-\frac{1}{2}\left\{ \sigma_{z}^{i}\sigma_{z}^{j},\rho\right\} \right)\nonumber \\
% &  & +N\sum_{i=1}^{N}\gamma(t)\left(\sigma_{z}^{a}\rho\sigma_{z}^{i}-\frac{1}{2}\left\{ \sigma_{z}^{i}\sigma_{z}^{a},\rho\right\} \right)\nonumber \\
% &  & +N\sum_{i=1}^{N}\gamma(t)\left(\sigma_{z}^{i}\rho\sigma_{z}^{a}-\frac{1}{2}\left\{ \sigma_{z}^{a}\sigma_{z}^{i},\rho\right\} \right)\nonumber \\
% &  & +N^{2}\gamma(t)\left(\sigma_{z}^{a}\rho\sigma_{z}^{a}-\rho\right)\}|2\rangle\nonumber \\
%= &  & [\left(iN\omega_{0}-\omega_{a}\right)-N^{2}\gamma(t)+N^{2}\gamma(t)\nonumber \\
% &  & +N^{2}\gamma(t)-N^{2}\gamma(t)]\rho_{1,2}(t)\nonumber \\
%= &  & \left(iN\omega_{0}-\omega_{a}\right)\rho_{1,2}(t),
%\end{eqnarray}
while the off-diagonal term is given by
\begin{equation}
\rho_{12}(t)=e^{-i\left(N\Omega_{0}-\omega_{a}\right)t}\rho_{12}(0).
\end{equation}

\begin{figure}
\includegraphics[bb=0 30 240 460,width=8.5cm]{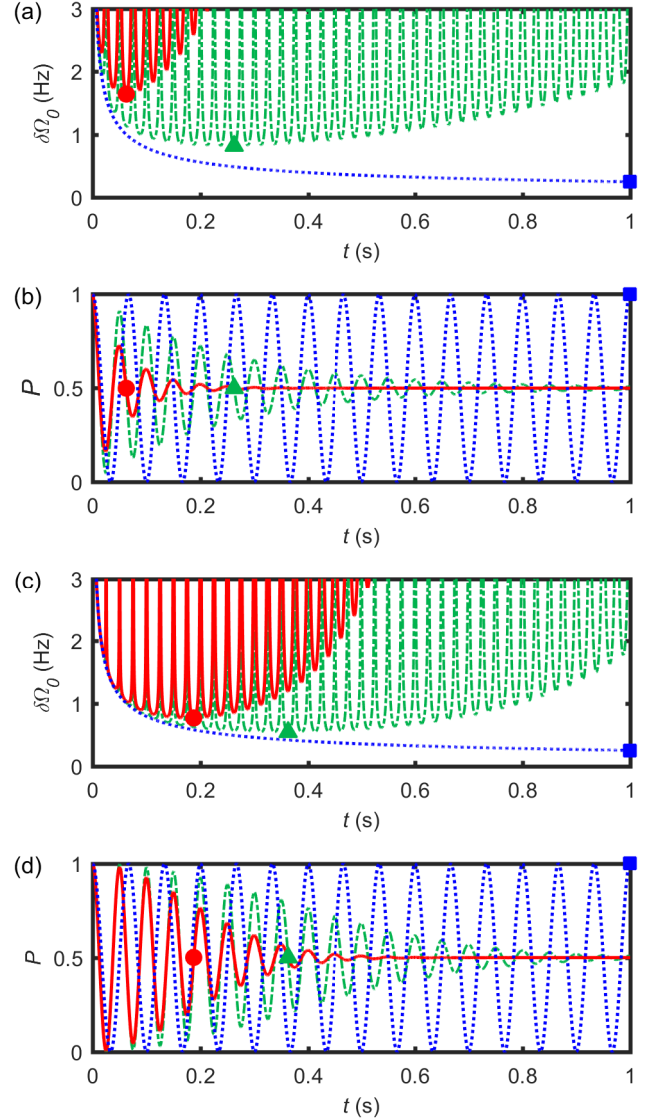}
\caption{The uncertainty of the measurement $\delta\Omega_0$ for (a) a Markovian dephasing dynamics
with $\Gamma(t)=\alpha t$, (c) a non-Markovian dynamics with $\Gamma(t)=\beta t^{2}$.
Equation~(\ref{eq:P}) for (b) a Markovian dephasing dynamics, (d) a non-Markovian dynamics.
The dashed green line indicates the uncorrelated decoherence. The solid red line indicates the superdecoherence in correlated environment. And the dotted blue line indicates our scheme.
The green triangle, red circle and blue square show the best interrogation time for uncorrelated decoherence, superdecoherence and our scheme, respectively. We use the following parameters, $n=4$, $\Omega_{0}=2\pi\times5$~Hz, $\alpha=1$~s$^{-1}$, $\beta=1$~s$^{-2}$, $T=1$~s.}\label{fig:dynamics}
\end{figure}

As in the conventional quantum metrology scheme, after a free evolution
for a time interval $t$ and a $\pi/2$ pulse, the probability of finding the probes in the
initial state reads
\begin{eqnarray}
P = \frac{1}{2}[1+\cos(N\Omega_{0}-\omega_{a})t].\label{eq:P}
\end{eqnarray}
We investigate Eq.~(\ref{eq:P}) for different decoherence dynamics in Fig.~\ref{fig:dynamics}. According to Eq.~(\ref{eq:deltaPhiE}), the entangled probes have a higher precision
when experiencing a longer evolution time $t_{e}$. Since the best interrogation time for the correlated bath is much shorter than that for the uncorrelated bath, the precision for the former is much worse. However,
in our scheme, the HL can be recovered because the effects of the noises on the working qubits have been effectively canceled due to the auxiliary qubit.
We further compare the case in non-Markovian dynamics with the one in Markovian dynamics. The best interrogation times are longer for the former, which is consistent with the prediction in Ref.~\cite{2012Quantum}.

%\begin{figure}
%\includegraphics[scale=0.9]{dynamics}
%\caption{ Equation~(\ref{eq:P}) for (a) a Markovian dephasing dynamics
%with $\Gamma(t)=\alpha t$, (b) a non-Markovian dynamics with $\Gamma(t)=\beta t^{2}$.
%The solid red line shows the probability for the uncorrelated decoherence.
%The dashed blue line shows the probability for the superdecoherence.
%And the dotted green line shows the probability for our scheme. The red circle and blue triangle show
%the best interrogation time for uncorrelated decoherence and superdecoherence,
%respectively. We use the following parameters, $n=10$, $\omega_{0}=2\pi\times10$~Hz,
%$\alpha=0.2~s^{-1}$ and $\beta=0.2~s^{-2}$. }\label{fig:dynamics}
%\end{figure}

According to Eq.~(\ref{eq:deltaPhiE}), the best resolution in the estimation %which was obtained with a given number of particles n and a total duration of experiment T,
satisfies $(N\Omega_{0}-\omega_{a})t_{e}=k\pi/2$ with odd $k$. The uncertainty in our scheme reads
%\begin{align}
%\delta\omega_{0}^{2} & =\frac{1}{\left(T/t\right)F\left(\omega_{0}\right)}\nonumber \\
% & =\frac{1-\cos^{2}\left[\left(N\omega_{0}-\omega_{a}\right)t\right]}{N^{2}Tt\sin^{2}\left[\left(N\omega_{0}-\omega_{a}\right)t\right]}.
%\end{align}

\begin{equation}
\delta\omega_{0}^{2}\mid_{e}=\frac{1}{N^{2}Tt_{e}}=\frac{1}{\left(n-1\right)^{2}Tt_{e}}\propto\frac{1}{\left(n-1\right)^{2}},\label{eq:SDe}
\end{equation}
where $n=N+1$. $N$ is smaller than $n$ due to the auxiliary qubit.

For unentangled probes, each probe needs % whose coupling constant satisfies $g_{k}^{a'}=g_{k}$ to let the dephasing rate to zero,
a properly-designed auxiliary qubit % $g_{k}^{a'}=g_{k}$
to cancel the effects of the noises on it. The uncertainty of the measurement is
\begin{equation}
\delta\omega_{0}^{2}\mid_{u}=\frac{1}{N^{\prime}Tt_{u}}=\frac{2}{nTt_{u}}\propto\frac{1}{n},
\end{equation}
where the best interrogation time $t_{u}$
is given by $(\Omega_{0}-\omega_{a})t_{u}=k\pi/2$ with odd
$k$. In this case, $n=2N^{\prime}$ and thus only half of the qubits play the role as the working qubits.

\begin{figure}
\includegraphics[bb=10 0 260 190,width=8.5cm]{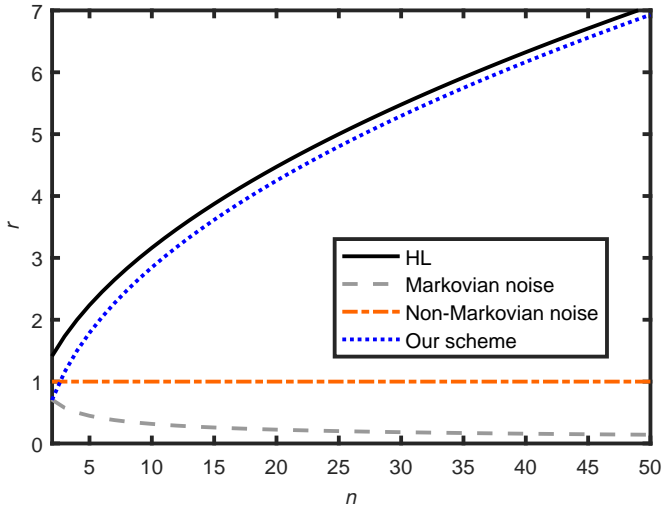}

\caption{\label{fig:The-relative-frequency}The relative ratio $r=\delta\phi|_u/\delta\phi|_e$ of the frequency resolutions
 of the entangled and unentangled probes when suffering from supercoherence.
The black solid line represents the noise-free HL. When there are no auxiliary qubits, the gray dashed line shows the case of Markovian noise, and the red dash-dotted line shows the case
of non-Markovian dynamics. And the blue dotted line shows our scheme
with only one auxiliary qubit.}
\end{figure}

Based on the quantum dynamics shown in Fig.~\ref{fig:dynamics}, we compare the uncertainty of measurement with and without auxiliary qubits
in the presence of supercoherence. In Fig.~\ref{fig:The-relative-frequency},
we consider two cases, i.e., the noise is fully Markovian with $\Gamma(t)\propto t$ or
non-Markovian dynamics with $\Gamma(t)\propto t^{2}$. When the noise is of non-Markovian dynamics, $r$ remains unity for different $n$'s. It implies that the precision of measurement can not be improved by using the entangled probes. When the noise is Markovian, $r$ approaches zero as $n$ increases, which suggests that using the entangled probes may even worsen the precision in the presence of superdecoherence. However, if one auxiliary qubit is employed to cancel the effects of noise, the precision in our scheme can approach the HL.

\subsubsection{Uncorrelated Environment}

When all probes are placed in an uncorrelated environment, e.g. spatially separated,
the correlation function becomes $C_{ij}(0)=\gamma(t)\delta_{ij}$. And the quantum master Eq.~(\ref{eq:SA equation}) is simplified as
\begin{eqnarray}
\dot{\rho} &=& i[\rho,H_\textrm{S}+H_\textrm{LS}]+\frac{1}{2}[\sum_i\gamma(t)(\sigma_{z}^{(i)}\rho\sigma_{z}^{(i)}-\rho)\nonumber \\
 &  & +C_{aa}(0,t)(\sigma_{z}^{a}\rho\sigma_{z}^{a}-\rho)].
\end{eqnarray}
Let $K=g_{k}^{a}/g_{k}$. Consequently, we have $C_{aa}(0,t)=K^{2}\gamma(t)$.
The off-diagonal element of the density matrix follows the evolution
\begin{eqnarray}
\rho_{12}(t)= \rho_{12}(0)e^{i(N\Omega_{0}-\omega_{a})t}e^{-(N+K^{2})\int\gamma(t^{\prime})dt^{\prime}}.
\end{eqnarray}
Since $N+K^{2}>N$, using auxiliary qubits can not reduce but increase the dephasing of probes. As a consequence, our measurement scheme can not improve the precision of measurement
in the case of uncorrelated environment.

\subsubsection{Partially-Correlated Environment}

In this subsection, we consider a partially-correlated environment. Assume that all
qubits, including $N$ working qubits and one auxiliary qubit, are spatially
arranged in a linear array, as shown in Fig.~\ref{fig:Schematic-illustration}. Following Ref.~\citep{PRA2013}, the real-valued homogeneous correlation functions $C_{ij}(0,t)$,
$C_{ia}(0,t)$ and $C_{ai}(0,t)$ read respectively
\begin{eqnarray}
C_{ij}(0,t)&=&e^{-x|i-j|}\gamma(t),\\
C_{ia}(0,t)&=&C_{ai}(0,t)=Ke^{-x(N+1-i)}\gamma(t),
\end{eqnarray}
where $x=d/\xi$ with $d$ being the spatial distance between two adjacent qubits, $\xi$
is the environmental correlation length \citep{PRA2013}. For an uncorrelated bath, we have $\xi=0$ and $C_{ij}=\delta_{ij}\gamma(t)$, while $\xi=\infty$ and $C_{ij}=\gamma(t)$ for any $i$ and $j$ in a fully-correlated bath. Generally, a finite but non-vanishing $\xi$ corresponds to a partially-correlated bath. The off-diagonal term of the density matrix is given by
\begin{eqnarray}
\rho_{12}(t)= \rho_{12}(0)e^{i(N\Omega_{0}-\omega_{a})t}e^{-A(N,x)\int\gamma(t^{\prime})dt^{\prime}}.
\end{eqnarray}
where the factor
\begin{eqnarray}
A(N,x)&=&(K-a)^2+b-a^2,\\
a&=&\sum_{i=1}^{n}\exp[-x(N+1-i)],\\
b&=&\sum_{i,j=1}^{n}\exp(-x|i-j|).
\end{eqnarray}
Here, $A(N,a)\gamma(t)$ represents the total dephasing rate of the working and auxiliary qubits. In order to improve the precision of measurement, we can always choose a proper $K$ to minimize the factor $A(N,x)$.
Therefore, employing auxiliary qubits can also improve the precision of a measurement for the entangled probes in a partially-correlated environment.

\subsection{Example for Magnetic-Field Sensing}
\label{subsection4}

As known to all, the direction and magnitude of the geomagnetic field are closely related to the position on the earth. Thus, various schemes have been put forward to measure the magnetic field in order to utilize geomagnetic for navigation, e.g., radical pair mechanism \cite{2013Biology} for avian navigation and the decoherence behaviors of the NV center in diamond \cite{Zhao2012,17APS,Jarmola2012,Yang2020,Head-Marsden2021}. In this section, as an example, we show how to utilize our measurement scheme with one auxiliary qubit in magnetic-field sensing. The initial
state of the entangled probes with $N$ working qubits and one auxiliary qubit is prepared in $|\Psi(0)\rangle=(|1\rangle_{a}|0\rangle^{\otimes N}+|0\rangle_{a}|1\rangle^{\otimes N})/\sqrt{2}$,
where $|0\rangle$ and $|1\rangle$ refer to the parallel and antiparallel
spin states with respect to the magnetic field, respectively. Then, let the probe be exposed to a magnetic field $B$. After a time interval $t$, it evolves into
\begin{eqnarray}
\rho(t)\!\! &=&\!\! \frac{1}{2}[|1\rangle_{a}\langle1|_{a}(|0\rangle\langle0|)^{\otimes N}+|0\rangle_{a}\langle0|_{a}(|1\rangle\langle1|)^{\otimes N}\nonumber \\
 \!\!&  & \!\! +e^{i(N\gamma_{0}B-\gamma_{a}B)t}e^{-\Gamma(t)}|1\rangle_{a}\langle0|_{a}(|0\rangle\langle1|)^{\otimes N}\nonumber \\
\!\! &  &\!\! +e^{-i(N\gamma_{0}B-\gamma_{a}B)t}e^{-\Gamma(t)}|0\rangle_{a}\langle1|_{a}(|1\rangle\langle0|)^{\otimes N}],
\end{eqnarray}
where $\gamma_{0}$ ($\gamma_{a}$) is the gyromagnetic ratio
of the working (auxiliary) qubit. %ÐýŽÅ±È¶šÒåÎªÔ­×ÓÔÚŽÅ³¡ÖÐœøÐÐÀ­Äª¶ûœø¶¯Ê±µÄœÇÆµÂÊÓëŽÅžÐÓŠÇ¿¶È$\gamma=\omega/B$
By choosing an appropriate
auxiliary qubit which satisfies $g_{k}^{a}=Ng_{k}$, we have
$\Gamma(t)=0$. The reduced density matrix of the working qubits at time $t$ is given by
\begin{eqnarray}
\rho(t) &=& \frac{1}{2}[|1\rangle_{a}\langle1|_{a}(|0\rangle\langle0|)^{\otimes N}+|0\rangle_{a}\langle0|_{a}(|1\rangle\langle1|)^{\otimes N}\nonumber \\
 &  & +e^{i(N\gamma_{0}B-\gamma_{a}B)t}|1\rangle_{a}\langle0|_{a}(|0\rangle\langle1|)^{\otimes N}\nonumber \\
 &  & +e^{-i(N\gamma_{0}B-\gamma_{a}B)t}|0\rangle_{a}\langle1|_{a}(|1\rangle\langle0|)^{\otimes N}].
\end{eqnarray}
After a $pi/2$ pulse, the probability of finding the probes in their initial state reads $P=[1+\cos(N\gamma_{0}Bt-\gamma_{a}Bt)]/2$.
Here, the Fisher information is given by
\begin{eqnarray}
F(B) = \frac{(N\gamma_{0}-\gamma_{a})^{2}t^{2}\sin^{2}(N\gamma_{0}Bt-\gamma_{a}Bt)}{1-\cos^{2}(N\gamma_{0}Bt-\gamma_{a}Bt)}.
\end{eqnarray}
Since the best interrogation time for the entangled probes is determined by $(N\gamma_{0}B-\gamma_{a}B)t_{e}=k\pi/2$
with odd $k$, we calculate the uncertainty
in the estimated value of $B$ as
\begin{equation}
\delta B^{2}|_{e}=\frac{1}{\left(N\gamma_{0}-\gamma_{a}\right)^{2}Tt}.\label{eq:B accuracy}
\end{equation}
When $N$ is large enough, the precision of measurement will approach the HL. %seeing from Eq.~(\ref{eq:B accuracy})

\section{Quantum Simulation of Quantum Metrology}
\label{Sec2}

In the NMR platform, we can use bath-engineering technique and GRAPE algorithm to simulate the quantum dynamics under superdecoherence.
By this quantum simulation, we can further prove that our measurement scheme with one auxiliary qubit can approach the HL, even in the case of superdecoherence.
Utilizing the bath-engineering technique, we apply a
time-dependent magnetic field to the total system including the working and auxiliary qubits. The total system is governed by the Hamiltonian \citep{WangEfficient,Nana2020}
\begin{equation}
H=\frac{1}{2}\Omega_{0}\sum_{i=1}^{N}\sigma_{z}^{(i)}+\frac{1}{2}\omega_{a}\sigma_{z}^{a}+\beta_{1}(t)\sum_{i=1}^{N}\sigma_{z}^{(i)}+\beta_{2}(t)\sigma_{z}^{a},\label{eq:simulation Hamiltonian}
\end{equation}
with
\begin{eqnarray}
\beta_{1}(t)&=&b_{1}\sum_{j=1}^{J}\omega_{j}F(j)\cos(\omega_{j}t+\psi_{1j}),\\
\beta_{2}(t)&=&b_{2}\sum_{j=1}^{J}\omega_{j}F(j)\cos(\omega_{j}t+\psi_{2j}),
\end{eqnarray}
where $b_{1}$ and $b_{2}$ are the noise amplitudes perceived
by the working and auxiliary qubits, respectively.
$\omega_{0}$ ($\omega_{J}=J\omega_{0}$) is the base (cutoff)
frequency with $\omega_{j}=j\omega_{0}$. $F(j)$ is the function which determines the
type of noise, and $\psi_{ij}$ ($i=1,2$) is a random number.
All the parameters above can be manipulated manually. Here, we apply identical time-dependent magnetic fields to all of the working qubits to simulate the superdecoherence. We further assume $b_{2}=nb_{1}$ and $\psi_{1j}=\psi_{2j}$ to cancel the effects of the noise by the auxiliary qubit.

We divide the Hamiltonian~(\ref{eq:simulation Hamiltonian}) into two parts,
i.e., the control Hamiltonian
$H_{c}=(\Omega_{0}\sum_{i=1}^{N}\sigma_{z}^{(i)}+\omega_{a}\sigma_{z}^{a})/2$,
and the noise Hamiltonian $H_{0}(t)=\beta_{1}(t)\sum_{i=1}^{N}\sigma_{z}^{(i)}+\beta_{2}(t)\sigma_{z}^{a}$.
In the Schr\"{o}dinger picture, the propagator of this dynamics is given by
\begin{eqnarray}
U(t)&=& U_{c}(t)\tilde{U}(t)
\end{eqnarray}
where
\begin{eqnarray}
U_{c}(t)&=& e^{-\frac{i}{2}(\Omega_{0}t\sum_{i}\sigma_{z}^{(i)}+\omega_{a}t\sigma_{z}^{a})},\nonumber \\
\tilde{U}(t) &=&  e^{-i\int_{0}^{t}d\tau\beta_{1}(\tau)\sum_{i}\sigma_{z}^{(i)}}e^{-i\int_{0}^{t}d\tau\beta_{2}(\tau)\sigma_{z}^{a}}.
\end{eqnarray}

The initial state of the total system is $|\Psi(0)\rangle=[|1\rangle_{a}|0\rangle^{\otimes N}+|0\rangle_{a}|1\rangle^{\otimes N}]/\sqrt{2}$. Let it evolve for a time interval $t$ under the Hamiltonian~(\ref{eq:simulation Hamiltonian}). Thus, we have
\begin{eqnarray}
|\psi(t)\rangle \!\!&=&\!\! U(t)|\psi(0)\rangle\nonumber \\
\!\!&=&\!\! \frac{1}{\sqrt{2}}(e^{-i\phi(t)}|1\rangle_{a}|0\rangle^{\otimes N}+e^{i\phi(t)}|0\rangle_{a}|1\rangle^{\otimes N}),
\end{eqnarray}
with
\begin{eqnarray}
\phi(t) &= & \phi_{A}(t)+\phi_{B}(t),\nonumber \\
\phi_{A}(t) &= & \frac{1}{2}(N\Omega_{0}-\omega_{a})t,\nonumber \\
\phi_{B}(t) &= & \frac{1}{2}[N\int_{0}^{t}d\tau\beta_{1}(\tau)-\int_{0}^{t}d\tau\beta_{2}(\tau)].
\end{eqnarray}
In order to mimic the effect of decoherence, we prepare a large number of ensemble which evolve under different Hamiltonians characterized by a set of the random numbers $\{\psi_{1j},\psi_{2j}\}$. Finally, the probability of finding the probes in the initial state is over the ensemble as
\begin{eqnarray}
P_{0}(t) \!\!=\!\! \frac{1}{2}[1+\cos2\phi_{A}\langle \cos2\phi_{B}\rangle  -\sin2\phi_{A}\langle \sin2\phi_{B}\rangle].
\end{eqnarray}

If we further assume a Gaussian noise, then we have $\langle \phi_{B}^{2m-1}(t)\rangle =0$
for any positive integer $m$ \citep{WangEfficient}, and thus yields
\begin{eqnarray}
P_{0}(t)
= \frac{1}{2}[1+\cos2\phi_{A}(t)e^{-2\chi(t)}],
\end{eqnarray}
where
\begin{eqnarray}
\chi(t)\!\!&=&\!\! \langle \phi_{B}^{2}(t)\rangle = \frac{4}{2\pi}\smallint_{-\infty}^{+\infty}\frac{d\omega}{\omega^{2}}S(\omega)\sin^{2}\frac{\omega t}{2},\nonumber\\
S(\omega)\!\!&=&\!\!\frac{1}{4}[N^{2}S_{11}(\omega)-NS_{12}(\omega)-NS_{21}(\omega)+S_{22}(\omega)].\nonumber
\end{eqnarray}
Here,
$S_{ij}(\omega)=\smallint_{-\infty}^{+\infty}dt\langle \beta_{i}(0)\beta_{j}(t)\rangle \exp(i\omega t)$ $(i,j=1,2)$
 is the Fourier transform of the two-time correlation function
$\langle \beta_{i}(t_{1})\beta_{j}(t_{2})\rangle $,
which depends on the time interval $t_{2}-t_{1}$.
And $S(\omega)$ is the total power spectral density of the noise, which describes the energy distribution of the stochastic signal in the frequency domain \citep{WangEfficient}.

We utilize the relation that $b_{2}=Nb_{1}$ and $\psi_{1j}=\psi_{2j}$
to obtain the two-time correlation functions as
\begin{eqnarray}
\langle \beta_{1}(t+\tau)\beta_{1}(t)\rangle \!\!&=&\!\!\frac{b_{1}^{2}\omega_{0}^{2}}{2}\sum_{j=1}^{J}j^{2}F(j)^{2}\cos(\omega_{j}\tau),\nonumber\\
\langle \beta_{1}(t+\tau)\beta_{2}(t)\rangle\!\!&=&\!\!\langle \beta_{2}(t+\tau)\beta_{1}(t)\rangle\nonumber\\  \!\!&=&\!\!\frac{Nb_{1}^{2}\omega_{0}^{2}}{4}\sum_{j=1}^{J}j^{2}F(j)^{2}\cos(\omega_{j}\tau),\nonumber\\
\langle \beta_{2}(t+\tau)\beta_{2}(t)\rangle \!\!&=&\!\!\frac{N^{2}b_{1}^{2}\omega_{0}^{2}}{2}\sum_{j=1}^{J}j^{2}F(j)^{2}\cos(\omega_{j}\tau).\nonumber
\end{eqnarray}
Thus, the power spectral densities are respectively given by
\begin{eqnarray}
S_{11}(\omega)&=&\frac{\pi b_{1}^{2}\omega_{0}^{2}}{2}\sum_{j=1}^{J}j^{2}F(j)^{2}[\delta(\omega-\omega_{j})+\delta(\omega+\omega_{j})],\nonumber\\
S_{12}(\omega)&=& S_{21}(\omega)\nonumber\\
&=& \frac{N\pi b_{1}^{2}\omega_{0}^{2}}{2}\sum_{j=1}^{J}j^{2}F(j)^{2}[\delta(\omega-\omega_{j})+\delta(\omega+\omega_{j})],\nonumber\\
S_{22}(\omega)\!\!&=&\!\!\frac{N^{2}\pi b_{1}^{2}\omega_{0}^{2}}{2}\sum_{j=1}^{J}j^{2}F(j)^{2}[\delta(\omega-\omega_{j})\!+\!\delta(\omega+\omega_{j})].\nonumber
\end{eqnarray}
Obviously, the total power spectral density reads
\begin{equation*}
S\left(\omega\right)=N^{2}S_{11}\left(\omega\right)-NS_{12}\left(\omega\right)-NS_{21}\left(\omega\right)+S_{22}\left(\omega\right)=0,
\end{equation*}
and the decoherence function also vanishes, i.e.,
\begin{equation*}
\chi(t)=\frac{4}{2\pi}\smallint_{-\infty}^{+\infty}\frac{d\omega}{\omega^{2}}S\left(\omega\right)\sin^{2}\frac{\omega t}{2}=0.
\end{equation*}

%Give a fixed total duration of experiment $T$, $T/t$ denotes the
%actual number of experiment using entangled probes, uncertainty calculated
%by the Fisher information,
We calculate the uncertainty of the measurement by the Fisher information as
%\begin{eqnarray}
%F\left(\omega_{0}\right)= &  & \frac{1}{P(1-P)}\left(\frac{\partial P}{\partial\omega_{0}}\right)^{2}\nonumber \\
% &  & \frac{N^{2}t^{2}\sin^{2}\left[\left(N\omega_{0}-\omega_{a}\right)t\right]}{1-\cos^{2}\left[\left(N\omega_{0}-\omega_{a}\right)t\right]},
%\end{eqnarray}

\begin{equation}
\delta\Omega_{0}^{2}=\frac{1}{\left(T/t\right)F\left(\phi\right)}=\frac{1-\cos^{2}\left[\left(N\omega_{0}-\omega_{a}\right)t\right]}{N^{2}Tt\sin^{2}\left[\left(N\omega_{0}-\omega_{a}\right)t\right]}.\label{eq:simulation accuracy}
\end{equation}
By similar derivation to Eq.~(\ref{eq:SDe}), we have $\delta\Omega_{0}^{2}\mid_{e}\propto\left(n-1\right)^{-2}$. We find the scaling law for our scheme can approach the HL when $N$ is large enough, even in the case of superdecoherence.

\begin{figure}
\includegraphics[bb=0 0 240 285,width=8.5cm]{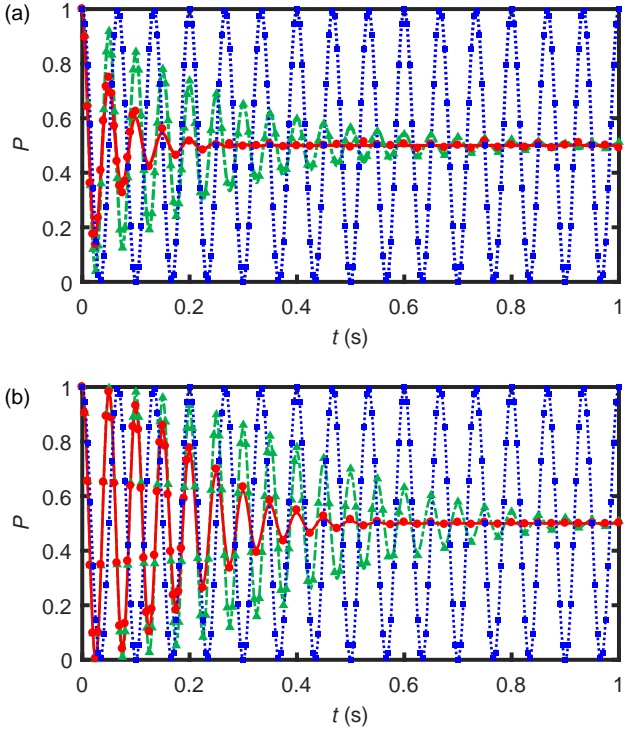}
\caption{Quantum simulation of population dynamics in Fig.~\ref{fig:dynamics}. The dashed green line, the solid red line and the dotted blue line are the same as Fig.~\ref{fig:dynamics}, while the green triangles, the red circle and the blue squares are obtained by the quantum simulation approach \cite{WangEfficient,Nana2020} for the uncorrelated decoherence, the superdecoherence and our scheme, respectively.}\label{Fig_RandomHamiltonian}
\end{figure}

\begin{figure}
\includegraphics[bb=0 0 260 200,width=8cm]{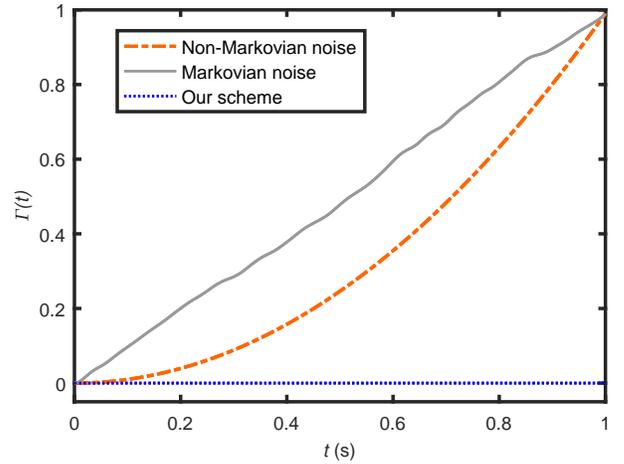}
\caption{The simulated decoherence factor $\Gamma(t)$ for different schemes: the gray solid line for Markovian environment, the red dash-dotted line for non-Markovian environment, and the blue dotted line for our scheme.}\label{Fig_DecoherenceFactor}
\end{figure}

We apply the quantum simulation algorithm to simulate our scheme as shown in Fig.~\ref{Fig_RandomHamiltonian}. We use the same
types of lines as Fig.~\ref{fig:dynamics} to represent the probability. Here, the number of random realizations in ensemble we used in this simulation is $2\times 10^3$. The time steps between the dots are 5~ms. Without loss of generality, we assume the model of white noise, i.e., $F(j)=1/j$ \cite{2014Soare}. The base and cutoff frequencies for the Markovian noise are $\omega_0=0.2$~Hz and $\omega_c=140$~Hz, respectively, while for the non-Markovian noise, we use the following parameters $\omega_0=10^{-3}$~Hz and $\omega_c=0.18$~Hz. We can learn from Fig.~\ref{Fig_RandomHamiltonian} that the results of stochastic Hamiltonian simulation are in quite-good agreement with the results of the Bloch-Redfield equation. The blue square dots represent our scheme, which implies that employing one auxiliary qubit can cancel the effects of noise. As a result, the precision in our scheme can approach the HL. In order to explore the underlying physical mechanism explicitly, we also plot the corresponding decoherence factor $\Gamma(t)$ in Fig.~\ref{Fig_DecoherenceFactor}. For the Markovian noise, the coherence decays with a constant rate, and thus using entangled probes will not improve the precision of the estimation. For the non-Markovian noise, since the coherence decays quadratically with time, the use of entangled probe can offer a better estimation, but the precision is still lower than the HL. However, for a correlated bath, because all qubits suffer from the collective noise, by properly arranging the auxiliary qubit the noises on the working qubits can be effectively canceled and thus the noise-free measurement can be performed.


\section{Conclusion}
In this paper, we propose a quantum metrology scheme with auxiliary states. The auxiliary states are properly designed and can selectively offset the impact of environmental noise. On account of superdecoherence, we analyze the optimal precision with and without auxiliary qubits. We find out that when the auxiliary qubits are not employed, the precision can not be improved by using the entangled probes no matter whether the noise is Markovian or non-Markovian. However, utilizing an auxiliary qubit can make the precision approach the HL in our scheme. We further discuss the cases of uncorrelated and partially-correlated environment, and find out that employing auxiliary qubits can improve the precision of measurement in the partially-correlated environment but it fails in the uncorrelated environment.
As an example, we show how to utilize our scheme with one auxiliary qubit in magnetic-field sensing. Finally, we use the bath-engineering technique and GRAPE to simulate the quantum dynamics and demonstrate that assisted by an auxiliary qubit our scheme can approach the HL in the case of superdecoherence.

%\section{Acknowledgment}
We thank valuable discussions with M.-J. Tao. This work is supported by the National Natural Science Foundation of China under Grant Nos.~11674033,~11474026,~11505007, and Beijing Natural Science Foundation under Grant No.~1202017.

\appendix*

\section{Dynamics of n-qubit Decoherence}

The quantum dynamics of open system is described by the well-known spin-boson
model \citep{Leggett1987Dynamics}. Since the time scale of dephasing
is generally much smaller than that of longitudinal relaxation,
we consider the pure-dephasing dynamics for simplicity.
The spin-boson Hamiltonian of $n$ qubits coupled to a common bath
can be written as \citep{2016Colloquium,PRA2013,2008QuantumJump,2014QuantumJump}
\begin{equation}
H=\frac{1}{2}\omega_{0}\sum_{i=1}^{n}\sigma_{z}^{(i)}+\sum_{k}\omega_{k}b_{k}^{\dagger}b_{k}+\sum_{i=1}^{n}\sigma_{z}^{(i)}\sum_{k}g_{k}^{(i)}(b_{k}^{\dagger}+b_{k}).\label{eq:Hamiltonian}
\end{equation}
Assuming that $\hbar=1$, the system Hamiltonian is $H_\textrm{S}=\omega_{0}\sum_{i=1}^{n}\sigma_{z}^{(i)}/2$,
and the bath Hamiltonian is $H_\textrm{B}=\sum_{k}\omega_{k}b_{k}^{\dagger}b_{k}$,
and the interaction Hamiltonian is $H_\textrm{I}=\sum_{i=1}^{n}\sigma_{z}^{(i)}\sum_{k}g_{k}^{(i)}(b_{k}^{\dagger}+b_{k})$,
where $\omega_{0}$ is the energy separation between the ground and
excited states, $b_{k}^{\dagger}$ ($b_{k}$) is the creation (annihilation)
bath operator, $g_{k}$ is the coupling constant between
the qubit and $k$th mode of bath, which is assumed to be real for simplicity.
%%%%%%%%%%%%%%%

The interaction Hamiltonian is the product of the system operators
and the bath operators, i.e., $H_\textrm{I}=\sum_{i=1}^{n}s_{i}B_{i}$ with $s_{i}=\sigma_{z}^{(i)}$ and
$B_{i}=\sum_{k}g_{k}^{(i)}(b_{k}^{\dagger}+b_{k})$. We decompose the system
operators $s_{i}$ into several parts in the eigenspace $\{|\epsilon\rangle_{i}\}$ as $s_{i}=\sum_\omega s_{i}(\omega)|\epsilon\rangle_{i}\langle\epsilon'|$ with $s_{i}(\omega)=\sum_{\epsilon'-\epsilon=\omega}\langle\epsilon|s_{i}|\epsilon'\rangle_{i}$.
The summation in $s_{i}(\omega)$ is extended over all energy eigenvalues $\epsilon$ and $\epsilon'$
of $H_{S}$ with a fixed energy difference of $\omega$ \citep{PRA2013,OpenQuantumSystems}.
Because $s_{i}=\sigma_{z}^{(i)}$, we have $s_{i}(\omega_{0})=s_{i}(-\omega_{0})=0$ and $s_{i}(0)=\mp1$ for $|0\rangle_{i}$ and $|1\rangle_{i}$, respectively.
Introducing these eigenoperator decompositions, the Bloch-Redfield
equations can be rewritten as
\begin{equation}
\dot{\rho}=i[\rho,H_\textrm{S}]+\sum_{i,j=1}^{n}D_{ij}(0,t)(\sigma_{z}^{(j)}\rho\sigma_{z}^{(i)}-\frac{1}{2}\{ \sigma_{z}^{(i)}\sigma_{z}^{(j)},\rho\} ),\label{eq:BR equation}
\end{equation}
where $D_{ij}(\omega,t)=\intop_{0}^{t}d\tau \exp(i\omega\tau)\langle\widetilde{B}_{i}(\tau)\widetilde{B}_{j}(0)\rangle$ %$D_{ij}(0,t)=\intop_{0}^{t}e^{i\omega\tau}d\tau\langle\widetilde{B}_{i}(\tau)\widetilde{B}_{j}(0)\rangle|_{\omega=0}=\intop_{0}^{t}d\tau\langle\widetilde{B}_{i}(\tau)\widetilde{B}_{j}(0)\rangle$
are the spectral functions, which define both temporal and spatial
correlations of the pure-dephasing noise environment \citep{JCP2015}.
$\widetilde{B}_{i}(\tau)=g_{k}^{(i)}[b_{k}^{\dagger}\exp(i\omega_k\tau)+b_{k}\exp(-i\omega_k\tau)]$
are the operators of environment in the interaction picture, and $\langle\widetilde{B}_{i}(\tau)\widetilde{B}_{j}(0)\rangle$
are the correlation functions, i.e.,
\begin{eqnarray}
  \langle \widetilde{B}_{i}(\tau)\widetilde{B}_{j}(0)\rangle = \sum_{k}g_{k}^{(i)}g_{k}^{(j)}[2\overline{n}(\omega_{k},T)\cos\omega_{k}\tau+e^{-i\omega_{k}\tau}].\nonumber\\
\end{eqnarray}
Thus, the spectral functions are explicitly given as
\begin{eqnarray}
D_{ij}(\omega,t) &=& \sum_{k}g_{k}^{(i)}g_{k}^{(j)}[2\overline{n}(\omega_{k},T)\frac{\sin\omega_{k}t}{\omega_{k}}\nonumber\\
&&+\frac{1-e^{-i\omega_{k}t}}{i\omega_{k}}]|_{\omega_{k}=\omega},
\end{eqnarray}
where $\overline{n}(\omega_{k},T)=\langle b_{k}^{\dagger}b_{k}\rangle $
denotes the average occupation number of mode $k$
at temperature $T$. The real part of $D_{ij}(0,t)$
causes dephasing while its imaginary part corresponds to Lamb shift \citep{PRA2013}. Let $D_{ij}(0,t)=\frac{1}{2}C_{ij}(0,t)+iF_{ij}(0,t)$, and $H_\textrm{LS}=\sum_{i,j=1}^{n}F_{ij}(0,t)\sigma_{z}^{(i)}\sigma_{z}^{(j)}$
is the Lamb-shift Hamiltonian. Thus, Eq.~(\ref{eq:BR equation})
is rewritten as
\begin{eqnarray}
\dot{\rho} &= & i[\rho,H_\textrm{S}+H_\textrm{LS}]\label{eq:BR new}\\
 &  & +\frac{1}{2}\sum_{i,j=1}^{n}C_{ij}(0,t)(\sigma_{z}^{(j)}\rho\sigma_{z}^{(i)}-\frac{1}{2}\{ \sigma_{z}^{(i)}\sigma_{z}^{(j)},\rho\} ).\nonumber
\end{eqnarray}
%with
%\begin{eqnarray}
%C_{ij}(\omega,t)&= & 2\mathrm{Re}[D_{ij}(\omega,t)]\\
%F_{ij}(\omega,t)&= &  \mathrm{Im}[D_{ij}(\omega,t)] .\label{eq:Fij}
%\end{eqnarray}

For $n=1$, since the total Hamiltonian is $H=\omega_{0}\sigma_{z}/2+\sum_{k}\omega_{k}b_{k}^{\dagger}b_{k}+\sigma_{z}\sum_{k}g_{k}(b_{k}^{\dagger}+b_{k})$,
Eq.~(\ref{eq:BR new}) is simplified as
\begin{equation}
\dot{\rho}=i[\rho,\frac{1}{2}\omega_{0}\sigma_{z}]+C(0,t)(\sigma_{z}\rho\sigma_{z}-\rho),
\end{equation}
where the single-qubit dephasing rate is given by $\gamma(t)=C(0,t)$.

However, the $n$-qubit dephasing rate varies with the environmental model.
For simplicity, we assume that $g_{k}^{(i)}=g_{k}^{(j)}=g_{k}$. When $n$ qubits are in
an uncorrelated environment, e.g. the distance between the qubits are
far beyond the correlation length of the environment \citep{PRA2013}, we can neglect all spatial correlations of the bath, i.e., $C_{ij}(0,t)=\gamma(t)\delta_{ij}$.
Hence $\gamma_{n}(t)=\sum_{i,j=1}^{n}C_{i,j}(0,t)=n\gamma(t)$. In contrast, when
$n$ qubits are in a correlated bath, i.e., the correlation length of the environment is much bigger than the qubits' spatial separation, the $n$-qubit dephasing
rate is $n^{2}$ times that of single qubit, i.e., $\gamma_{n}(t)=n^{2}\gamma(t)$, resulting from $C_{ij}(0,t)=\gamma(t)$ for all $i$ and $j$. This phenomenon is called superdecoherence,
mainly due to collective entanglement between qubits and environment \citep{Palma1996Quantum}.

%\begin{figure}
%\includegraphics{figureRho12}
%\caption{(a) a Markovian dephasing dynamics
%with $\Gamma(t)=\alpha t$, (b) a non-Markovian dynamics with $\Gamma(t)=\beta t^{2}$.
%The solid red line shows the quantum dynamics for the superdecoherence.
%The dashed blue line shows the quantum dynamics for the uncorrelated decoherence.
%We use the following parameters, $n=10$, $\omega_{0}=2\pi\times10$~Hz,
%$\alpha=0.2~s^{-1}$ and $\beta=0.2~s^{-2}$. }\label{fig:dynamics of dephasing}
%\end{figure}

\bibliography{reference}

\end{document}